# BERT4Loc: BERT for Location - POI Recommender System

**Syed Raza Bashir [1], Shaina Raza [2], and Vojislav Misic [3]**

[1] syedraza.bashir@torontomu.ca
[2] shaina.raza@vectorinstitute.ai
[2] vojislav.misic@torontomu.ca

**Abstract:** Recommending points of interest (POIs) is a challenging task that requires extracting comprehensive location data from location-based social media platforms. To provide effective location-based recommendations, it's important to analyze users' historical behavior and preferences. In this study, we present a sophisticated location-aware recommendation system that uses Bidirectional Encoder Representations from Transformers (BERT) to offer personalized location-based suggestions. Our model combines location information and user preferences to provide more relevant recommendations compared to models that predict the next POI in a sequence. Our experiments on two benchmark dataset show that our BERT-based model outperforms various state-of-the-art sequential models. Moreover, we see the effectiveness of the proposed model for quality through additional experiments.

## 1. Introduction

The popularity of mobile communication and the internet is leading us towards an era of mobile networks. This popularity has also made it easier to determine the geographic location of users through smart devices like mobile phones and iPads. As a result, location-based social networks (LBSNs) are becoming increasingly popular, with well-known sites like Foursquare [1] and Gowalla [2]. Users can check-in at any location and time on major social networking sites, rate or comment on Points of Interest (POIs), and create their social network to exchange POIs with others who share similar interests.

The rapid increase in mobile internet users and the volume of available information has created a conflict between limited network resources and growing user demands. To address this challenge, there is a need to explore algorithms that can accurately and efficiently extract users' behavioral patterns, analyze their interests, and provide personalized POI recommendations. Recommender systems [3,4] are specialized software that analyze users' preferences, behaviors, and interactions to generate personalized suggestions, such as POIs in our context. By incorporating a recommender system, we can optimize network resources, address the challenges posed by the growing information volume, and enhance user satisfaction.

Developing a POI recommendation system based on LBSNs is a challenging, compared to building standard recommendation algorithms. This is due to the multifaceted contextual data involved in POI recommendations, including users' check-in time, social network, POI category information, geographical location, and more. Moreover, users' interests can shift over time and vary across locations. Another significant challenge is data sparsity, which arises from the limited number of POIs visited by users. To offer personalized POI recommendations, it is important to effectively integrate POIs and model the data derived from them.

Deep neural network-based models [5] have achieved state-of-the-art success in various domains and applications, including social networks, health science, news, and entertainment. Among these models, Transformer-based [6] architectures are the most promising. These architectures employ self-attention mechanisms [6] to differentially weight the significance of each component in the input data. In this study, we propose a POI recommendation model based on Bidirectional Encoder Representations from Transformers (BERT) [7], an advanced neural-network based model developed by Google researchers in 2018.

BERT has demonstrated exceptional accuracy in a range of natural language processing (NLP) tasks, such as general language understanding, classification, summarization, and translation [8]. The model is trained simultaneously on both Masked Language Modeling (MLM) (50%) and Next Sentence Prediction (NSP) (50%) tasks. MLM enables bidirectional learning from text by masking a word in a sentence and leveraging BERT's predictive capability to identify the hidden word. NSP helps BERT learn sentence relationships by predicting whether a given sentence follows the previous one or not.

For our POI recommendation system, we adapt the BERT model, specifically repurposing the MLM and NSP tasks for a location-aware context. We chose BERT due to its extensive training on vast data sources like Wikipedia and the Toronto Books Corpus, enabling it to predict "masked" items (e.g., POIs in our research). Moreover, BERT effectively handles sequential data and time-ordered user sequences with high accuracy [9,10]. In our context, sequential modeling allows us to predict users' next actions based on their historical preferences.

The steps of our work are summarized as follows: (1) We prepare users' histories from the dataset by arranging each user's history as a time-sorted list of POIs; (2) We replace some of these POIs with a [MASK] token, using the MLM task; (3) We train our model on POI datasets, such as Yelp and Foursquare, to predict the correct values of the masked POIs. Through this process, our model learns the useful representations that exist between different POIs. As our recommendation model is based on the BERT architecture, we name it BERT4Loc (Bidirectional Encoder Representations from Transformers for Location). The specific contributions of this paper are manifold, which are given below:

1) We adapt BERT, a state-of-the-art NLP model, to the location-aware recommender system domain, resulting in the proposed BERT4Loc model.

2) BERT4Loc effectively captures users' sequential interactions and fully utilizes location information for recommending POIs to users.

3) We demonstrate the applicability and effectiveness of BERT4Loc on real-world datasets, showing improved performance compared to existing location-aware recommender systems.

The novelty of our work lies in adapting the BERT model to the context of location-aware recommender systems and successfully implementing it to improve POI recommendations. There are some state-of-the-art recommender systems that have been proposed to address the challenges associated with POI recommendations. Some notable methods include matrix factorization-based techniques [11]; recurrent neural networks (RNN) [12], adversarial models [13]. Matrix factorization methods, such as GeoMF [11] incorporate geographical information to improve recommendations, while RNN-based models, like ST-RNN [12] exploit the temporal dynamics of users' check-in behavior. Adversarial-based models, such as Geo-ALM [13] capture the complex interactions between users and locations through a fusion of geographic features and generative adversarial networks.

Despite the effectiveness of previous methods, there are certain limitations when it comes to capturing the complex relationships between users and POIs or dealing with data sparsity. First, previous methods struggle to effectively capture the complex relationships between users and POIs. Second, these methods can also face challenges in dealing with data sparsity, which occurs when there are limited POI visits by users. To address these limitations, our proposed BERT4Loc model offers a promising solution by leveraging the advanced capabilities of BERT in handling sequential data and contextual relationships. By applying BERT in location-aware recommender systems, we can overcome these limitations and provide more accurate and personalized POI recommendations.

The rest of the paper is organized as follows: Section 2 describes related work. Section 3 presents the proposed methodology and Section 4 describes the experimental setup. Section 5 discusses the results and analysis. Finally, Section 6 concludes the paper.

## 2. Related Work

This section covers some state-of-the-art works in this line of research.

The recommender systems assume that the users who share behavioral preferences are more likely to form connections with each other. Social recommender systems target the social media domain [14] such as Facebook, Instagram, Twitter. SocialMF [15] is a social recommender system that learns the users' preferences using the basic matrix decomposition model and then builds the profile by using the information of user's friends. SoReg [16] is another social recommender that solves the overfitting problem in SocialML. TrustSVD [17], another social recommender system, models the implicit behavioral data and social relationship of users. By incorporating implicit information about the user's network, this model brings the user's feature representation closer to actual application scenarios.

In 2016, Google proposed the Wide&Deep [5] approach, which has arisen a lot of interest in the recommendation systems. This model has inspired several other models, including DeepFM [18], Deep & Cross [19], NFM (Neural Factorization Machines) [20], AFM (Attentional Factorization Machines) [21], DIN (Deep Interest

Network) [22], and others. Sequential recommender systems [23], based on deep neural networks, also model the real-life interactions between users' actions before and after they have consumed an item. For example, in a location recommendation station, when a person visits a place, it is more appropriate to recommend other POIs that are nearby and related to the user's current and past visited POIs.

With recent progress in NLP, sequential recommendation models have also advanced. From the initial Markov Chain [24], through the following RNNs, Convolutional Neural Network (CNN) models, and now the popular Transformer [6,25], there has been a lot of progress. The SASRec [26] sequence recommendation model is based on the self-attention mechanism to model past user behavior and retrieve top-k recommendations.

Building on top of the Wide&Deep [27], researchers developed the Behavior Sequence Transformer (BST) [27] model. The BST model follows the popular embedding and Multi-Layer Perceptron (MLP) paradigm, where the previous click item and related features are first embedded into a low-dimensional vector and then entered into the MLP. The most significant difference between BST and WDL is the addition of a Transformer layer [22]. The Transformer layer helps to learn better representations of items that the user has clicked on by capturing the underlying sequential signals. Currently, BST is being used in the Rank stage of Taobao recommendation, providing recommendation services to hundreds of millions of consumers daily. BERT4Rec [10] is another Transformer-based recommender model that interprets the sequence of user actions as if it were a text sequence. However, this model lacks user modeling and contextual information.

In this work, we use the BERT model for the POI recommendations. We include more item (POI) information, and we also include more user's features including their social relationships. Our model is trained on a real-world dataset and can be used in real-time for POI recommendations.

## 3. Methodology

We formalize the problem and introduce our training objectives in this section. The notations used in this table are given in Table 1.

**Table 1.** Notations Used

| Notation | Description |
|---|---|
| $U$ | Set of users |
| $V$ | Set of items (POIs) |
| $S_u$ | List of interactions of user $u$ with items |
| $n_u$ | Number of interactions of user $u$ |
| $v_t^u$ | Item at the relative time step $t$ for user $u$ |
| $K_v$ | Set of keywords describing item $v$ |
| $K$ | Set of side (metadata) information related to the items |
| $K^*$ | Set of all possible keyword combinations |
| $E_V$ | Embedding of the POI (item) identifier |
| $E_P$ | Embedding for the position of items in the sequence |
| $N$ | Input sequence length |
| $h_t^0$ | Sum of item embedding $e_t$ and the positional embedding $p_t$ |
| $k_t$ | Embedding of the keywords $K_{v_t}$ of item $v_t$ |
| $L$ | Number of Transformer layers |
| $h_t^L$ | Last hidden state of the $L_{th}$ Transformer layer |
| $BPR$ | Bayesian Personalized Ranking |
| $X$ | Number of sampled negative items in uniform distribution |

We address the problem of recommending POI items to users based on their previous interactions. Let $U = \{u_1, u_2, \ldots, u_{|U|}\}$ be a set of users and $V = \{v_1, v_2, \ldots, v_{|V|}\}$ be a set of items. The list of interactions of user $u \in U$ is denoted by $S_u = \{v_1^u, v_2^u, \ldots, v_n^u\}$, where user $u$ has interacted with item $v_t^u \in V$ at the relative time step $t$. Each item $v \in V$ is associated with a set of keywords $K_v = \{k_1, k_2 \ldots, k_{|K_v|}\}$ that describe the item $v$. We refer to set $K$ as the metadata related to the item. Given the history $S_u$ and the additional metadata $K_{v_t}^{u*}$ for every $v_t^u \in S_u$, the POI recommendation task is to predict the next item $v_{n_{u+1}}^{u}$ in the sequence of user's interactions. We adapted the deep bidirectional self-attention model BERT [7] for the sequential recommendation task, resulting in our BERT4Loc model. The modified BERT4Loc architecture is shown in Figure 1 and consists of three different layers:

- Embedding layer: This layer learns a representation of the inputs, including the POI (business) ID and the associated metadata (e.g., business category). The output of the embedding layer is fed to the next layer, the Transformer layer.
- Transformer layer: This layer comprises L layers of Transformer blocks; we use 12 layers and 12 attention heads. Each *layer* takes in a list of token embeddings and produces the same number of embeddings on the output (with transformed feature values). The output of the final Transformer block is passed to the projection layer.
- Projection layer: This layer projects the learned hidden representation from the previous layer into the item space for prediction using a softmax layer. We employ the Cloze task *[28]* for training, where the model predicts randomly masked items in the interaction sequence.

To incorporate metadata, which consists of item descriptions, we make modifications to the original BERT4Rec model, which are discussed below:

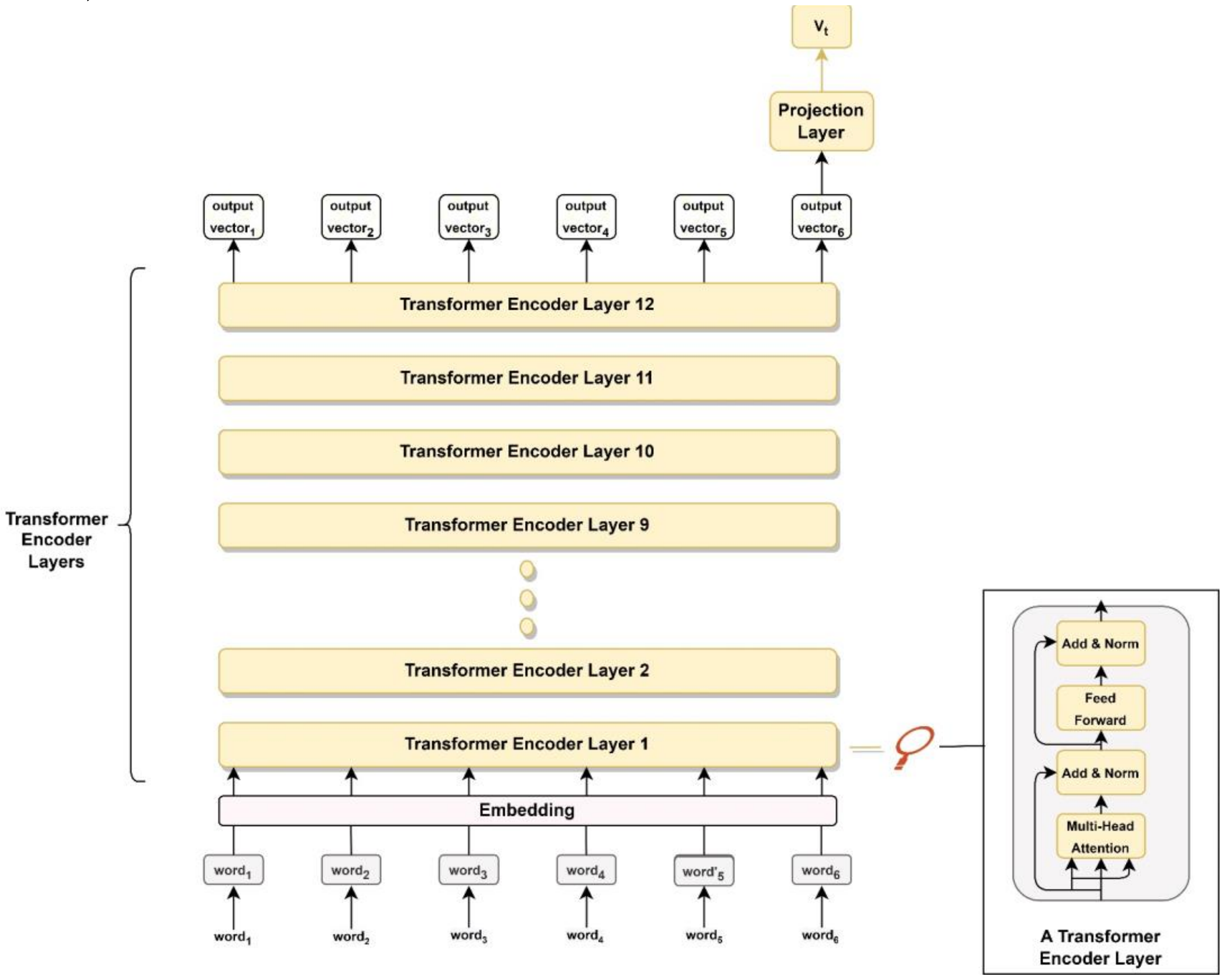

**Figure 1.** BERT4Loc Architecture

*Modification in the Transformer layer*: As the BERT4Rec model has two embeddings: (i) an embedding $E_V \in R^{(|V|\times d)}$ of the POI (item) identifier, and (ii) an additional embedding $E_P \in R^{N\times d}$ for the position of the items in the sequence for the Transformer blocks, where $N$ is the input sequence length, and the input to the Transformer layer is the sum of the item embedding $e_t$ and the positional embedding $p_t$ for every sequence step t, denoted as $h_t^0 = e_t + p_t$. Our first modification to the BERT4Rec is that we modify this Transformer layer and add an additional embedding $k_t$ of the keywords $K_{v_t}$ of item $v_t$ representing the metadata as an additional summand: $h_t^0 = e_t + p_t + k_t$. To accomplish this, we merge all keywords of item $v_t$ into a super keyword $K_{v_t}^{u*}$ and embed it using $E_{K*} \in R^{(|K*|\times d)}$ where $K^*$ is the set of all possible keyword combinations. We then encode the categories as a multi-hot vector, scaled to the embedding size d using a linear layer. The keyword descriptions are masked accordingly during training and evaluation.

*Modification during training*: The last hidden state in BERT4Rec is the $L_{th}$ Transformer layer, $h_t^L$ of the masked item $v_t$ at time step $t$, is used as a linear layer with GELU as the activation function. This projection layer is the final part of the model to decide which item to recommend to the user based on their preferences and interaction history. The model predicts users' preference scores for POI items and recommends the top-scoring POI items in the prediction process.

Our second modification is that during the training process, we employ Bayesian Personalized Ranking (BPR) to calculate the loss (which is different from the BERT4Rec). We predict users' preference scores for POI items and recommend the top-scoring POI items during the prediction process. We utilize a sample-based ranking approach [29,30], where each positive item is paired with $X$ sampled negative items, uniformly distributed.

## 4. Experimental Setup

In this section, we elaborate the experiment setup.

### *4.1. Data Set*

We utilized the Yelp Dataset [31], which is available on Yelp's website. The dataset contains 1.6 million reviews and 500,000 tips from 366,000 users for 61,000 businesses. It also includes 481,000 business attributes, such as hours, parking availability, and ambiance, as well as check-ins for each of the 61,000 businesses collected over time via a social network of 366,000 people, resulting in a total of 2.9 million social edges. Specifically, the dataset comprises data for 61,184 businesses, 1,569,264 reviews, and 366,715 users.

We also used the Foursquare Dataset [1] to address the recommendation problem. From this dataset, we obtained approximately 43,108 unique geographical locations for our experiment. We considered 18,107 users with a total of 2,073,740 check-ins, focusing on users with at least 10 check-ins.

For user features from both datasets, we considered user ID, user reviews, ratings, and timestamps of user interactions. For location (item) features, we used location ID, location name, and city name. The user ID and business ID serve as primary information related to users and businesses (locations), respectively. Metadata related to users includes stars, text (review), and interaction timestamps, while metadata related to items consists of business names. We included these pieces of information for training, but our model is generalizable and can accommodate more metadata if available.

The features of both the datasets used in this work are given in Table 2.

**Table 2.** Datasets features used.

| **Dataset** | **Unique Locations** | **Users** | **Check-ins** | **Minimum Check-ins per User** | **Features (User)** | **Features (Locations)** |
|---|---|---|---|---|---|---|
| Yelp | 61,184 | 366,715 | 1,569,264 | N/A | User ID, User Reviews, Ratings, Timestamps | Location ID, Business Name, Category. |
| Foursquare | 43,108 | 18,107 | 2,073,740 | 10 | User ID, User Reviews, Ratings, Timestamps | Location ID, Location Name, Category. |

In this work, we converted all numeric ratings or the presence of a review to implicit feedback of 1 (i.e., the user interacted with the item). We then grouped interaction records by users and constructed interaction sequences for each user by sorting the records by timestamps.

We also perform some analysis of the data. Figure 2 calculates the distribution of ratings and computes the average length of reviews in Yelp dataset. We observe in Figure 2 shows that most of the reviews have lengths of 100, 200, and 400 , which is likely due to the fact that these are common values for people to use when writing reviews. Many people may have a certain amount of information they want to convey in their review and may find that these lengths allow them to do so effectively.

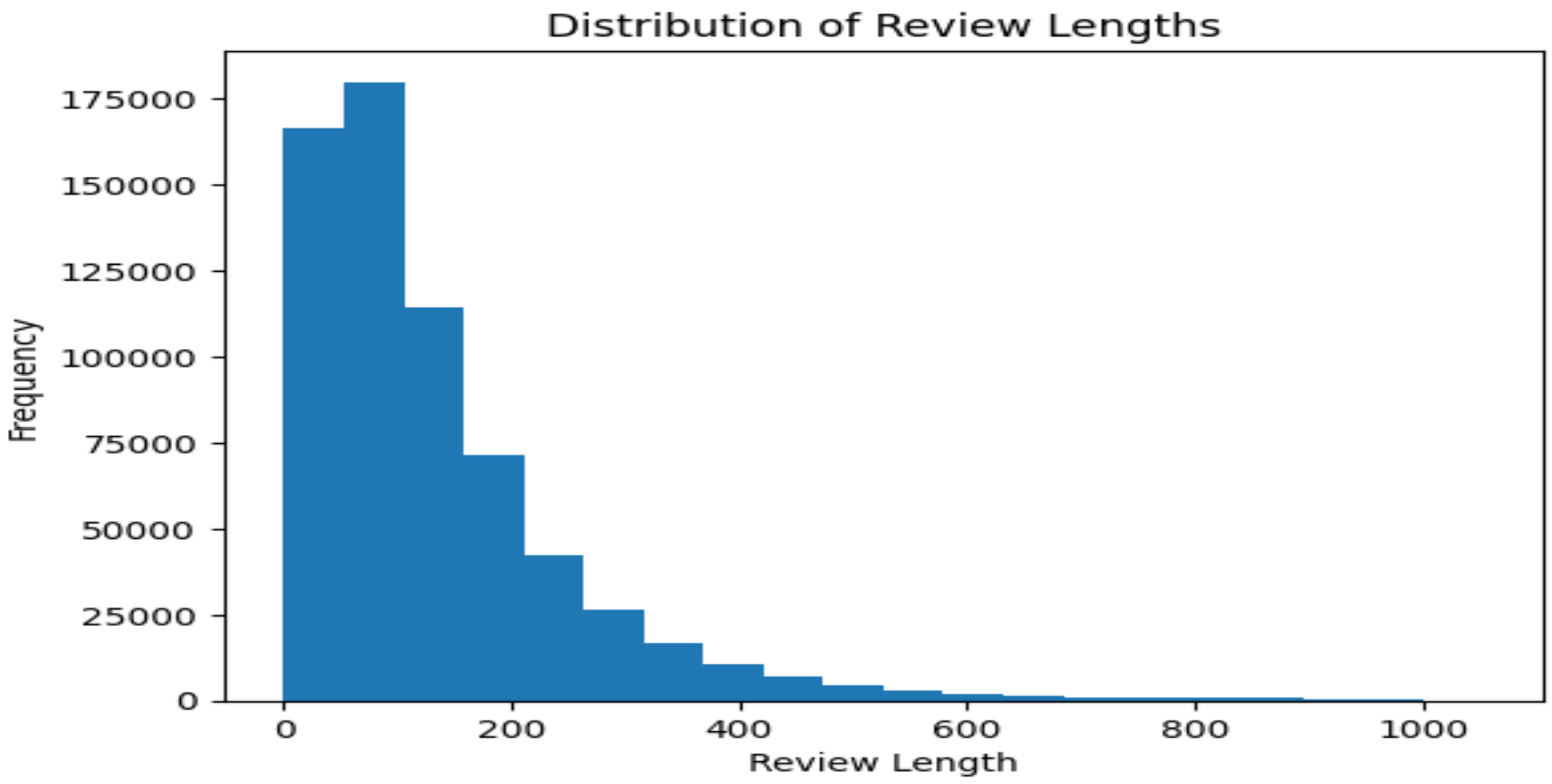

**Figure 2.** Chart showing the distribution of ratings in the Yelp Review dataset.

Next, we show the word cloud of positive reviews in both dataset in Figure 3. The word cloud of positive reviews shows that some of the most common words used in positive reviews are "food", "place", "great", "good", and "nice one". These are, in general, words that people commonly use to describe positive experiences with restaurants and other food-related businesses.

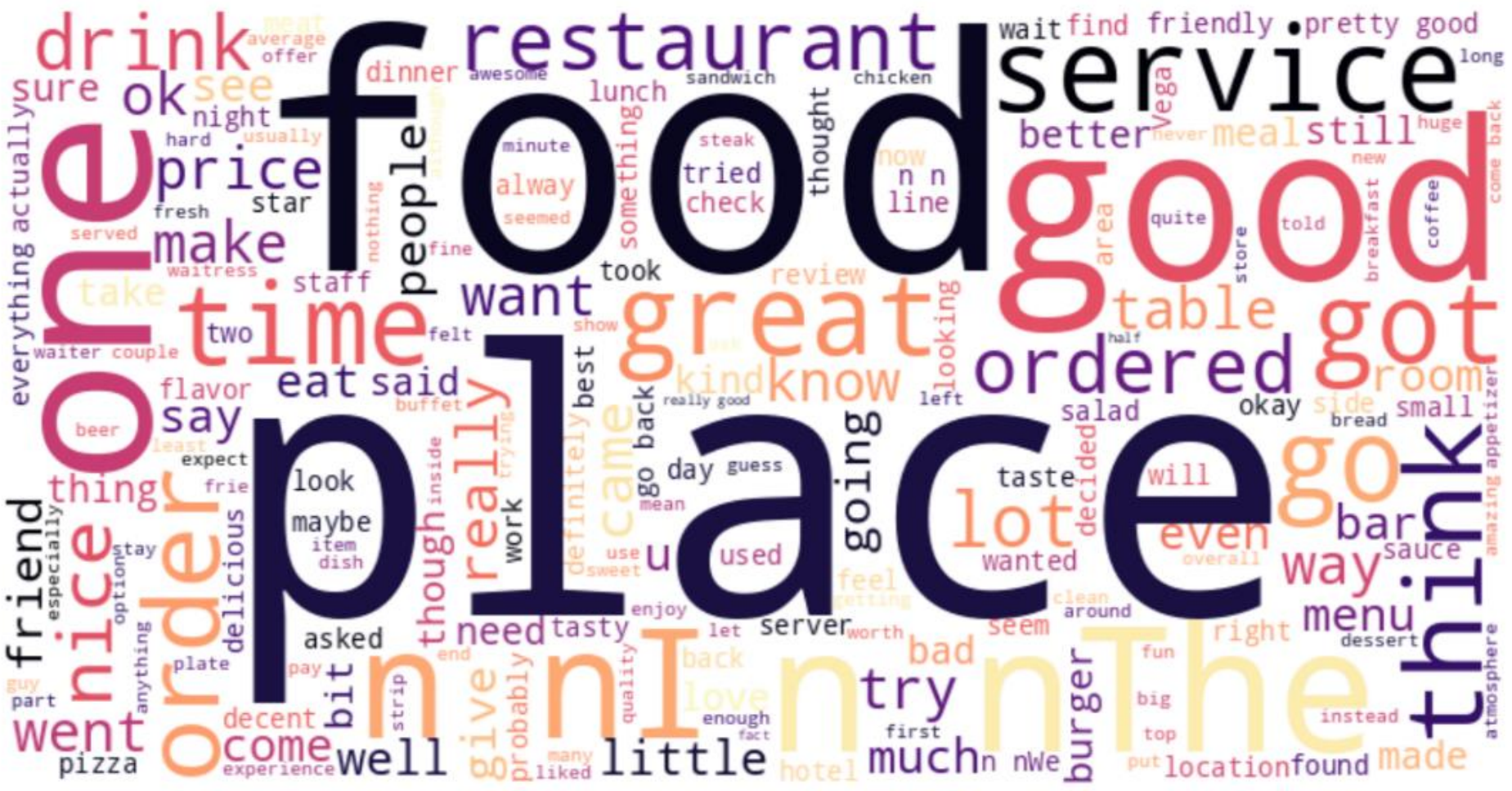

**Figure 3.** Word cloud of positive reviews in the both datasets.

We also performed the word cloud for all the negative reviews in the dataset (i.e., those with a rating of 1 or 2) and do not find any useful patterns. So, we performed the sentiment analysis using VADER [32] library performed on the reviews data from both the dataset and reported the distribution of positive, negative, and neutral sentiments on reviews, which ranges from -1 (most negative) to 1 (most positive) in Figure 4.

Based on the sentiment distribution histogram in Figure 4, we can observe that there are some negative sentiments expressed by reviewers, but the majority of the sentiment polarities fall between neutral and 75% positive.

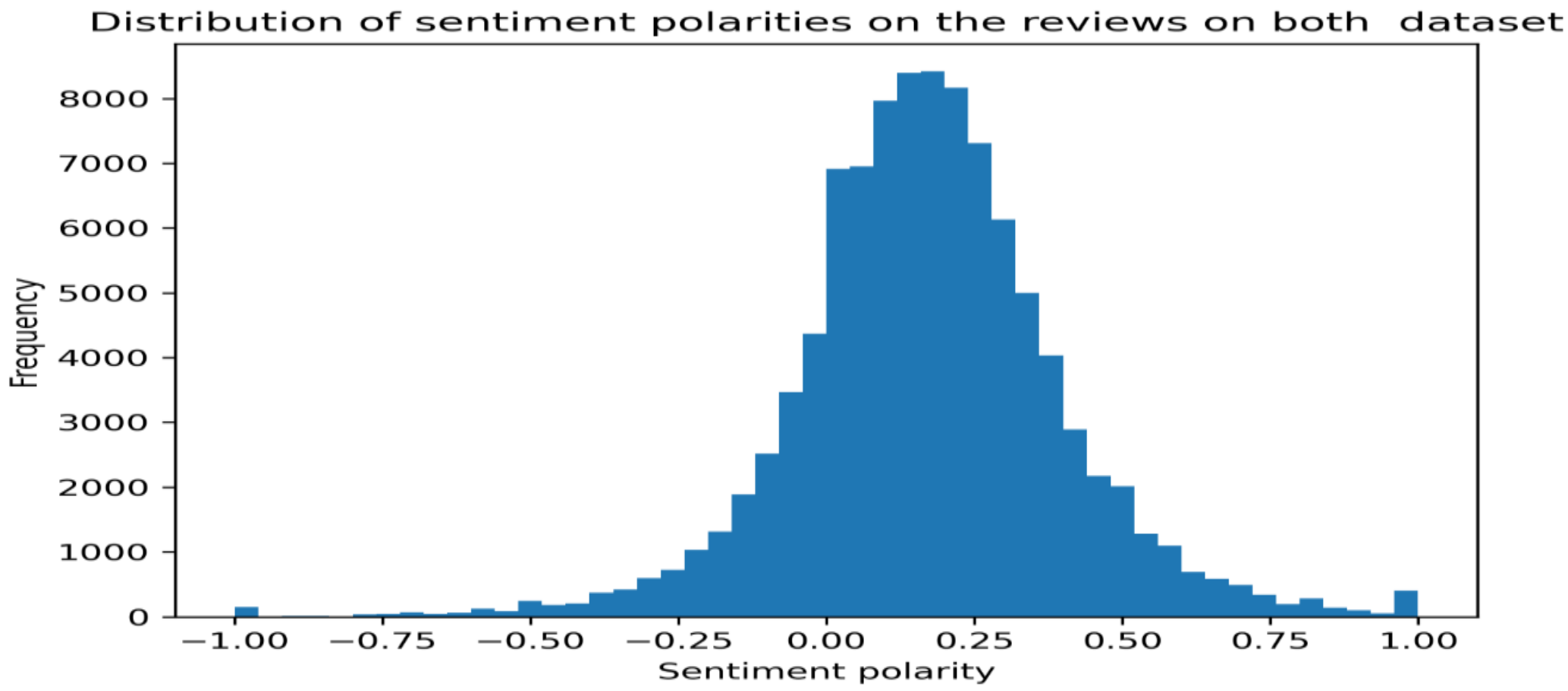

**Figure 4.** Distribution of sentiment polarities on the reviews of both datasets.

Figure 5 displays the distribution of reviews and star ratings across major cities in Yelp dataset. Our analysis found that Las Vegas had the highest number of ratings. However, when looking at the distribution of 5-star ratings for different cities (as shown in Figure 5), they are almost equal.

| city | review_count | stars |
|---|---|---|
| Las Vegas | 161325.0 | 3.710267 |
| Phoenix | 58722.0 | 3.657419 |
| Toronto | 42528.0 | 3.475424 |
| Scottsdale | 26709.0 | 3.933251 |
| Charlotte | 25480.0 | 3.554627 |
| Pittsburgh | 18290.0 | 3.602564 |
| Henderson | 16665.0 | 3.808628 |
| Tempe | 15605.0 | 3.754717 |

**Figure 5.** Cities with the most reviews and best ratings.

### *4.2. Evaluation Methodology and Metric*

For our experiments, we used a time-based splitting criterion on both datasets. In time-based splitting, the test set indexes are greater than those of the validation set, which are greater than those of the training set. As an example, if a dataset contains at least twenty ratings for each item and user, the available ratings can be divided into train, validation, and test sets. The test set contains the user's most recent six ratings, while the validation set contains the user's previous four ratings. The train set for this dataset is comprised of all previous ratings (at least ten).

In our work, we sorted the data by the timestamp of user interaction in descending order and split them accordingly. We used the leave-one-out data splitting method where, for each user, we held out the last item of the behavior sequence as the test data, treated the item just before the last as the validation set, and utilized the remaining items for training. We use the following evaluation metrics in this paper.

*Precision* is a measure for computing the fraction of relevant items out of all the recommended items. We average the metric for each user u to get the final result. Precision is shown in Equation (1).

$$Precision@k = \frac{1}{|U|}\sum_{u \in U} \frac{|\hat{R}(u) \cap R(u)|}{|\hat{R}(u)|} \quad (1)$$

Where $R_u$ represents the ground-truth set of items that user $u$ has interacted with and $|\hat{R}(u)|$ represents the item count of $\hat{R}(u)$.

*Recall* is a measure for computing the fraction of relevant items out of all relevant items. Recall is defined in Equation (2).

$$Recall@k = \frac{1}{|U|}\sum_{u \in U} \frac{|\hat{R}(u) \cap R(u)|}{|R(u)|} \quad (2)$$

*Hit-Ratio (HR)* is a way of calculating how many 'hits' are in a k sized list of ranked items. If there is at least one item that falls in the ground-truth set, it is called a hit. HR is defined in Equation (3).

$$HR@k = \frac{1}{|U|}\sum_{u \in U} \delta(\hat{R}(u) \cap R(u) \neq \emptyset) \quad (3)$$

Where δ(.) is an indicator function, δ (.)=1 if it is a hit and 0 otherwise. ∅ denotes the empty set.

*Normalized Discounted Cumulative Gain (NDCG)* is a measure of ranking quality, where positions are discounted logarithmically. It accounts for the position of the hit by assigning higher scores to hits at top ranks. It is defined as in Equation (4).

$$NDCG@k = \frac{1}{|U|}\sum_{u \in U} \left(\frac{1}{\sum_{i=1}^{\min(|R(u)|,K)} \frac{1}{log_2(i+1)}} \sum_{i=1}^{K} \delta(i \in R(u)) \frac{1}{log_2(i+1)}\right) \quad (4)$$

where δ(.) is an indicator function. We consider the values of top @k, where k is 10, 20 and 50. We compared the performance of various recommender systems on top@k metrics, including Precision, Recall, F1-score, Hit Rate (HR), and Normalized Discounted Cumulative Gain (NDCG), with their respective Mean and Standard Deviation (Mean±SD) values. The results were reported on both datasets. Regarding the calculation of Mean values, we took the average of the performance metric values across all users or items in the test set. For example, to calculate the Mean Precision, we first calculated the Precision value for each user or item in the test set, and then took the average of these values. The Standard Deviation (SD) was calculated to measure the variability of the performance metric values across all users or items in the test set.

### *4.3. Baselines*

To evaluate the performance of our models, we compared them with a range of different approaches to recommendation systems. These baselines were chosen based on the most commonly used methods in recommender system research. By comparing our models to these baselines, we were able to assess their effectiveness and identify areas for improvement.

These baseline are given below:

*BERT4Rec* [10]: Sequential Recommandation with Bidirectional Encoder Representations from Transformer, which employs the deep bidirectional self-attention to model user behavior sequences.

*MultiVAE* [33]: Variational Autoencoders for collaborative filtering extends the variational autoencoders (VAEs) for implicit feedback. This model uses a non-sampling method by default, so we also did not use any negative sampling in this method. We simply use cross-entropy loss for this method as the loss function type.

*ENMF* [34]: Efficient Neural Matrix Factorization without sampling for recommendation is based on a matrix factorization architecture that learns from the whole training data to make recommendations. We use the cross-entropy as the loss function for this model.

SASRecF [35]: Feature-level Deeper Self-Attention Network for sequential recommendation that integrates heterogeneous features of items into feature sequences with different weights through an attention mechanism. We chose the BPR as the loss function.

*RepeatNet* [36]: Repeat Aware Neural Recommendation Machine for session-based recommendation that uses an encoder-decoder architecture to address repeat consumption in the session-based recommendation task. We chose the BPR as the loss function.

*SLIM* [37]: Sparse Linear Methods for Top-N recommendations introduces a linear model that learns to predict the similarity between items in a sparse manner. This method provides a compact and interpretable model for top-*N* recommendations. We use the BPR loss function for this model.

*NCF* [38]: Neural Collaborative Filtering is a general framework for collaborative filtering that combines matrix factorization and a multi-layer perceptron to learn the user-item interaction patterns. For this model, we use the cross-entropy loss function.

*GRU4Rec* [24]: Session-based Recommendations with RNNs - Gated Recurrent Units (GRU) models the sequential behavior of users in session-based recommendations. We use the BPR loss function for training.

*FPMC* [39]: Factorizing Personalized Markov Chains for next-basket recommendation combines matrix factorization and Markov chains to model user behavior and generate personalized recommendations. We use the BPR loss function for this model.

### *4.4. Hyperparameters setting*

In our experimental setup, we fine-tuned our BERT4Loc model with a set of carefully chosen hyperparameters. The loss function used in training was BPR, and we trained the model for 20 epochs with a learning rate of 0.001. We evaluated the model performance on top-k recommendations, where k was set to [10, 20, 50]. The evaluation metrics used were Recall, Precision, F1-score, HR, and NDCG. We used a train_batch_size and eval_batch_size of 16 and set the attention dropout probability to 0.5. Additionally, we set the mask_ratio to 0.2 to reduce overfitting.

We experimented with different hyperparameters to optimize the model's performance. We varied the number of layers in the user and item encoders, the number of attention heads, the size of the hidden layers, and the maximum sequence length. After a comprehensive search, we found that the optimal number of layers for both the user and item encoders was three, with six attention heads and a hidden layer size of 256. The maximum sequence length was set to 100, which was the maximum number of POIs visited by a user in our dataset.

To ensure a fair comparison, we fine-tuned all the baseline models to their optimal hyperparameter settings. We used Google Colab as the programming platform and connected it to Google Drive to store and access our data. During training, we created small batches of data to fit all the data into memory. Each batch took approximately 10-15 minutes for training, and the complete training process with the full dataset took several hours.

## 5. Results and Analysis

In this section, we present the results and analyze the performance of our model. we also test the effectiveness of the proposed approach for the sampling methods, recommendations techniques and report the results.

### *5.1. Overall Results*

The comparison between our model and all the baselines is shown in Table 3.

**Table 3.** Performance comparison of various recommender systems on top@k metrics, including Precision, Recall, F1-score, Hit Rate (HR), and Normalized Discounted Cumulative Gain (NDCG), with their respective Mean and Standard Deviation (Mean±SD) values. The results are reported on both datasets.

| Model | top@k | Precision (Mean±SD) | Recall (Mean±SD) | F1-score (Mean±SD) | HR (Mean±SD) | NDCG (Mean±SD) |
|---|---|---|---|---|---|---|
| **Yelp Dataset** | | | | | | |
| **BERT4Loc** | 10 | 0.56±0.04 | 0.45±0.05 | 0.50±0.05 | 0.82±0.06 | 0.42±0.03 |
| | 20 | 0.52±0.03 | 0.60±0.07 | 0.56±0.05 | 0.91±0.04 | 0.51±0.03 |
| | 50 | 0.49±0.02 | **0.78±0.09** | **0.60±0.06** | **0.92±0.03** | **0.71±0.03** |
| **BERT4Rec** | 10 | **0.61±0.04** | 0.43±0.05 | 0.50±0.05 | 0.65±0.05 | 0.52±0.01 |
| | 20 | 0.57±0.03 | 0.49±0.06 | 0.54±0.06 | 0.72±0.04 | 0.67±0.03 |
| | 50 | 0.43±0.03 | 0.72±0.08 | 0.54±0.06 | 0.86±0.05 | 0.70±0.03 |
| **MultiVAE** | 10 | 0.28±0.03 | 0.27±0.04 | 0.27±0.03 | 0.63±0.05 | 0.33±0.02 |
| | 20 | 0.23±0.02 | 0.39±0.05 | 0.29±0.03 | 0.78±0.06 | 0.32±0.03 |
| | 50 | 0.18±0.02 | 0.48±0.07 | 0.26±0.03 | 0.87±0.05 | 0.33±0.03 |
| **RepeatNet** | 10 | 0.26±0.02 | 0.13±0.03 | 0.17±0.02 | 0.20±0.03 | 0.18±0.02 |
| | 20 | 0.21±0.01 | 0.24±0.04 | 0.22±0.03 | 0.23±0.03 | 0.20±0.02 |
| | 50 | 0.18±0.01 | 0.41±0.06 | 0.25±0.03 | 0.34±0.04 | 0.24±0.03 |
| **SASRecF** | 10 | 0.22±0.02 | 0.14±0.03 | 0.17±0.02 | 0.12±0.02 | 0.12±0.01 |
| | 20 | 0.20±0.02 | 0.18±0.04 | 0.19±0.03 | 0.16±0.03 | 0.15±0.02 |
| | 50 | 0.19±0.01 | 0.22±0.04 | 0.20±0.02 | 0.22±0.03 | 0.16±0.02 |
| **ENMF** | 10 | 0.12±0.01 | 0.14±0.03 | 0.13±0.02 | 0.20±0.0 | 0.17±0.02 |
| | 20 | 0.11±0.01 | 0.19±0.03 | 0.14±0.02 | 0.19±0.03 | 0.18±0.02 |
| | 50 | 0.10±0.01 | 0.20±0.04 | 0.13±0.02 | 0.24±0.03 | 0.20±0.02 |
| **SLIM** | 10 | 0.33±0.03 | 0.32±0.04 | 0.32±0.03 | 0.55±0.05 | 0.29±0.02 |
| | 20 | 0.28±0.02 | 0.40±0.05 | 0.33±0.03 | 0.68±0.06 | 0.31±0.03 |
| | 50 | 0.23±0.02 | 0.55±0.07 | 0.32±0.03 | 0.80±0.05 | 0.35±0.03 |
| **NCF** | 10 | 0.41±0.04 | 0.35±0.05 | 0.38±0.04 | 0.64±0.06 | 0.25±0.02 |
| | 20 | 0.35±0.03 | 0.45±0.06 | 0.39±0.04 | 0.76±0.05 | 0.30±0.03 |
| | 50 | 0.29±0.02 | 0.65±0.08 | 0.40±0.05 | 0.86±0.04 | 0.38±0.03 |
| **GRU4Rec** | 10 | 0.39±0.03 | 0.27±0.04 | 0.32±0.03 | 0.61±0.05 | 0.28±0.02 |
| | 20 | 0.34±0.02 | 0.38±0.05 | 0.36±0.03 | 0.74±0.05 | 0.33±0.03 |
| | 50 | 0.27±0.02 | 0.54±0.07 | 0.36±0.04 | 0.82±0.05 | 0.37±0.03 |
| **FPMC** | 10 | 0.30±0.03 | 0.21±0.03 | 0.25±0.02 | 0.47±0.04 | 0.22±0.02 |
| | 20 | 0.25±0.02 | 0.29±0.04 | 0.27±0.03 | 0.58±0.05 | 0.24±0.02 |
| | 50 | 0.20±0.01 | 0.37±0.05 | 0.26±0.03 | 0.68±0.06 | 0.27±0.04 |

**Foursquare Dataset**

| | | | | | | |
|---|---|---|---|---|---|---|
| **BERT4Loc** | **10** | **0.54±0.03** | **0.42±0.04** | **0.48±0.04** | **0.81±0.05** | **0.40±0.03** |
| | 20 | 0.50±0.03 | 0.58±0.06 | 0.54±0.04 | 0.89±0.04 | 0.49±0.03 |
| | 50 | 0.47±0.02 | **0.76±0.08** | **0.58±0.05** | **0.91±0.03** | **0.69±0.03** |
| **BERT4REC** | 10 | **0.59±0.03** | 0.40±0.04 | 0.48±0.04 | 0.63±0.05 | 0.50±0.01 |
| | 20 | 0.55±0.03 | 0.47±0.05 | 0.51±0.03 | 0.70±0.04 | 0.65±0.02 |
| | 50 | 0.41±0.02 | 0.70±0.07 | 0.52±0.05 | 0.84±0.05 | 0.68±0.03 |
| **MultiVAE** | 10 | 0.27±0.03 | 0.25±0.03 | 0.26±0.02 | 0.61±0.05 | 0.32±0.02 |
| | 20 | 0.22±0.02 | 0.37±0.04 | 0.28±0.02 | 0.76±0.05 | 0.31±0.03 |
| | 50 | 0.17±0.02 | 0.46±0.06 | 0.25±0.02 | 0.85±0.04 | 0.32±0.03 |
| **RepeatNet** | 10 | 0.25±0.02 | 0.12±0.02 | 0.16±0.02 | 0.19±0.03 | 0.17±0.02 |
| | 20 | 0.20±0.01 | 0.23±0.03 | 0.21±0.02 | 0.22±0.03 | 0.19±0.02 |
| | 50 | 0.17±0.01 | 0.39±0.05 | 0.24±0.02 | 0.33±0.04 | 0.23±0.03 |
| **SASRecF** | 10 | 0.21±0.02 | 0.13±0.02 | 0.16±0.02 | 0.11±0.02 | 0.11±0.01 |
| | 20 | 0.19±0.02 | 0.17±0.03 | 0.18±0.02 | 0.15±0.03 | 0.14±0.02 |
| | 50 | 0.18±0.01 | 0.21±0.03 | 0.19±0.02 | 0.21±0.03 | 0.15±0.02 |
| **ENMF** | 10 | 0.11±0.01 | 0.13±0.02 | 0.12±0.02 | 0.19±0.03 | 0.16±0.02 |
| | 20 | 0.10±0.01 | 0.18±0.03 | 0.13±0.02 | 0.18±0.03 | 0.17±0.02 |
| | 50 | 0.09±0.01 | 0.19±0.03 | 0.12±0.02 | 0.23±0.03 | 0.19±0.02 |
| **SLIM** | 10 | 0.32±0.03 | 0.30±0.03 | 0.31±0.02 | 0.54±0.05 | 0.28±0.02 |
| | 20 | 0.27±0.02 | 0.38±0.04 | 0.32±0.02 | 0.67±0.05 | 0.30±0.03 |
| | 50 | 0.22±0.02 | 0.53±0.06 | 0.31±0.02 | 0.79±0.04 | 0.34±0.03 |
| **NCF** | 10 | 0.39±0.03 | 0.33±0.04 | 0.36±0.03 | 0.63±0.05 | 0.24±0.02 |
| | 20 | 0.34±0.03 | 0.43±0.05 | 0.38±0.03 | 0.75±0.04 | 0.29±0.03 |
| | 50 | 0.28±0.02 | 0.63±0.07 | 0.39±0.04 | 0.85±0.03 | 0.37±0.03 |
| **GRU4Rec** | 10 | 0.38±0.03 | 0.26±0.03 | 0.31±0.02 | 0.60±0.04 | 0.27±0.02 |
| | 20 | 0.33±0.02 | 0.36±0.04 | 0.34±0.02 | 0.73±0.04 | 0.32±0.03 |
| | 50 | 0.26±0.02 | 0.52±0.06 | 0.35±0.03 | 0.81±0.04 | 0.36±0.03 |
| **FPMC** | 10 | 0.29±0.02 | 0.20±0.03 | 0.24±0.02 | 0.46±0.03 | 0.21±0.02 |
| | 20 | 0.24±0.02 | 0.28±0.03 | 0.26±0.02 | 0.57±0.04 | 0.23±0.02 |
| | 50 | 0.19±0.01 | 0.36±0.04 | 0.25±0.02 | 0.67±0.03 | 0.26±0.03 |

Table 3 presents the performance of our approach and different recommender systems on various top@k metrics for both datasets. Based on the results, BERT4Loc and BERT4REC show the highest precision, recall, and F1-score among all models. These models demonstrate a strong performance across all top@k values (10, 20, and 50). BERT4Loc achieves the highest HR and NDCG for all top@k values, indicating that it is highly effective in ranking relevant items for users. The comparison between BERT4Rec and BERT4Loc reveals that BERT4Loc performs better in location-aware recommendation scenarios. While BERT4Rec is designed to take in user-item features without considering contextual features, BERT4Loc is able to leverage rich contextual features to make more accurate recommendations. Although our experiments involve giving the same input features to all baseline models, a model's ability to naturally incorporate rich contextual features can give it an advantage in providing better recommendations.

MultiVAE, RepeatNet, SASRecF, and ENMF models show relatively lower performance compared to BERT4Loc and BERT4REC. Their precision, recall, and F1-score values are lower, indicating that these models are less accurate in predicting the top-N items for users. The HR and NDCG metrics also indicate that these models are less effective in ranking relevant items.

SLIM, NCF, GRU4Rec, and FPMC models show moderate performance in comparison to the other models. Their precision, recall, and F1-score values are better than MultiVAE, RepeatNet, SASRecF, and ENMF but lower than BERT4Loc and BERT4REC. The HR and NDCG metrics for these models are also intermediate, indicating that they have moderate effectiveness in ranking relevant items.

The superiority of the our BERT-based model is attributed to its design, which has the following distinctive attributes: 1) it considers the side information from the POI items in the item encoder; 2) it considers the contexts in the user model to better capture the sequential correlation of a user’s POI history; and 3) it is based on sequential recommendation, which implicitly consider both the short and long-term interests of users.

Inherently, our model is based on the BERT architecture, which is a deep bidirectional self-attention model that can capture item relations on both sides, left and right, whereas other sequential recommendation models (in these experiments) only consider users’ historical sequences from left to right. In real situations, a user’s behavior depends on the user’s current interests, which can evolve in a highly dynamic manner. Therefore, considering only previous items is insufficient in terms of accuracy. Our model can predict the next item by extracting user historical patterns based on the relationship between rated items in the history data of target users (item sequences).

To summarize, in both datasets, BERT4Loc and BERT4REC demonstrate the highest precision, recall, and F1-score among all models, indicating their strong performance across different data settings. The HR and NDCG metrics also show that BERT4Loc performs well in ranking relevant items for users in both datasets. For the remaining models, their performance may vary between the two datasets due to the differences in data characteristics. Some models might be more sensitive to the data sparsity or the distribution of user-item interactions, which could lead to variations in their performance metrics.

Due to similar patterns in the results and for brevity reasons, we will report the results of the next experiments solely on the Yelp dataset.

### *5.2.Effectiveness of Different Sampling Techniques*

We test the effectiveness of different sampling techniques that are mostly used in recommender systems [30,40] in this experiment. We try the following sampling methods:

- *Full ranking:* evaluating the model on all set of items.
- *Uniform X (uni-X):* uniformly sample X negative items for each positive item in the testing set, and evaluate the model on these positive items with their sampled negative items.

- *Popularity X (pop-X):* sample X negative items for each positive item in the testing set based on item popularity, and evaluate the model on these positive items with their sampled negative items.

The results for this experiment are shown in Figure 6. In our initial experiments, we tried different values of uni-X and pop-X and found 100 to be a better value for both X. To provide a high-level summary of these results, we take the average of top @k (10, 20, and 50) to show the precision, recall, HR, NDCG, and F1 scores.

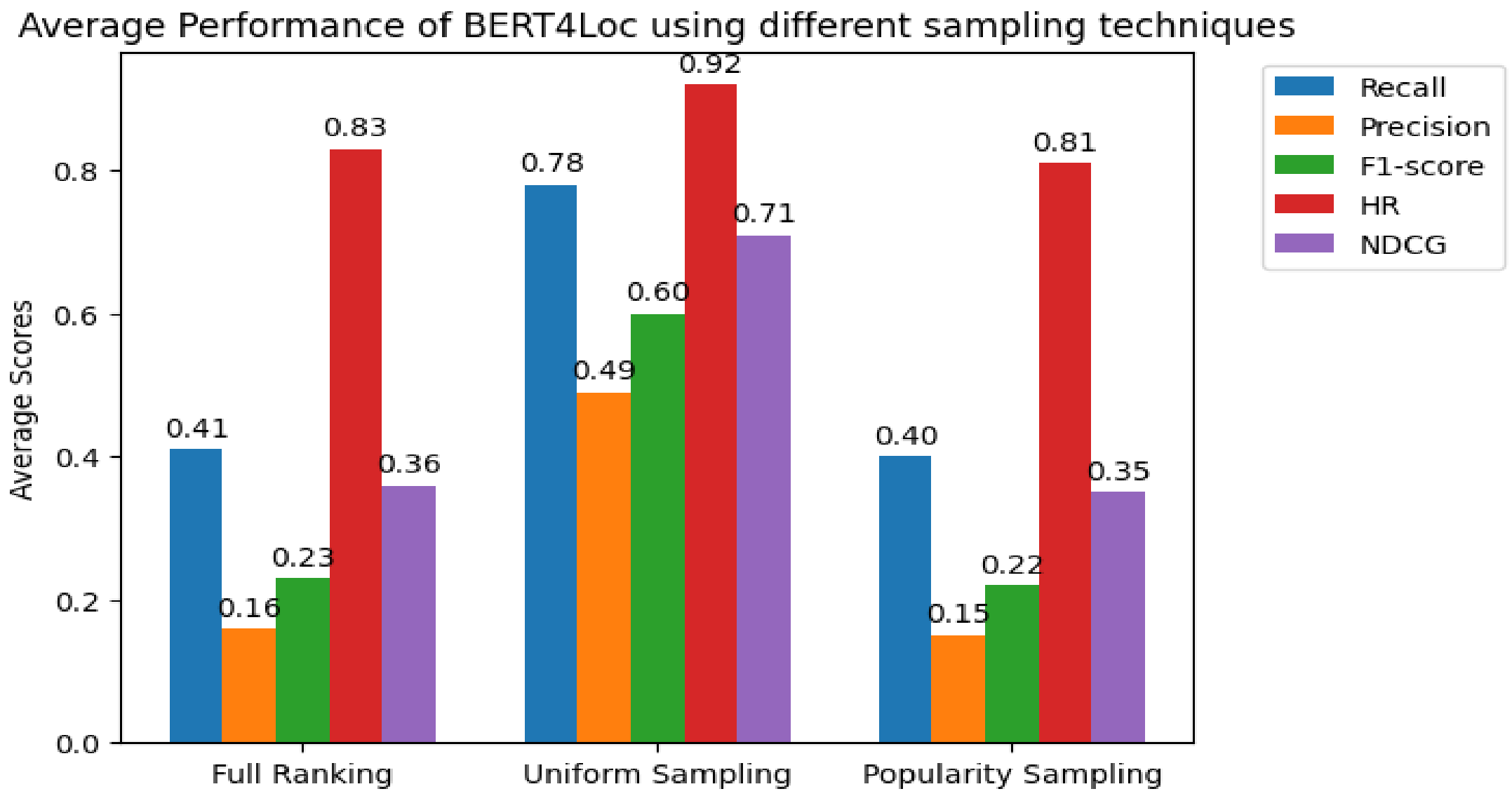

**Figure 6.** Average performance of BERT4Loc using different sampling techniques.

Figure 6 shows the average performance of BERT4Loc using different sampling techniques: Full Ranking, Uniform Sampling (uni-100), and Popularity Sampling (pop-100). According to this, Uniform Sampling (uni-100) consistently outperforms both Full Ranking and Popularity Sampling (pop-100) in terms of average recall, precision, F1-score, HR, and NDCG. This suggests that Uniform Sampling (uni-100) is the most effective sampling technique among the three.

Full Ranking exhibits a moderate performance compared to the Uniform sampling technique. It has higher average scores (though marginal) than Popularity Sampling (pop-100) but lower than Uniform Sampling (uni-100). This indicates that although Full Ranking provides relatively accurate recommendations, it may not be as effective as Uniform Sampling (uni-100).

Popularity Sampling (pop-100) has the lowest average scores across all metrics, indicating that it is the least effective sampling technique in this experiment. This is because it samples negative items based on item popularity, which may lead to some random recommendations and lower accuracy scores.

Overall, we find that given the dataset size and computational resources, using the negative sampling technique (Uniform Sampling) proves to be more effective than both Full Ranking and Popularity Sampling for the BERT4Loc.

### *5.3. Effectiveness of the Length of the Recommendation List*

We assess the recommendation accuracy for the top-k (k = [10, 100]) using F1-score (harmonic mean of precision and recall), HR and NDCG in Fig. 7.

The plot in Figure 7 shows the effectiveness of the length of the recommendation list for top-k (k = [10, 100]) using F1-score (harmonic mean of precision and recall), HR, and NDCG. We can observe that the model's accuracy improves as the value of k increases. However, after a certain point (k=50), the accuracy of the recommendations begins to decrease. This decrease in accuracy can be attributed to the fact that there may not be enough relevant items after a certain threshold of recommendations.

As the recommendation list becomes longer, it becomes more challenging for the model to provide accurate and relevant recommendations. Consequently, the precision and recall values start to decline, which in turn leads to a decrease in the F1-score, HR, and NDCG scores. Therefore, it is crucial to find an optimal value of k that balances the trade-off between the number of recommendations and the accuracy of the model. Based on the plot, it seems that k=50 is an optimal point where the accuracy is maximized before it starts to decline.

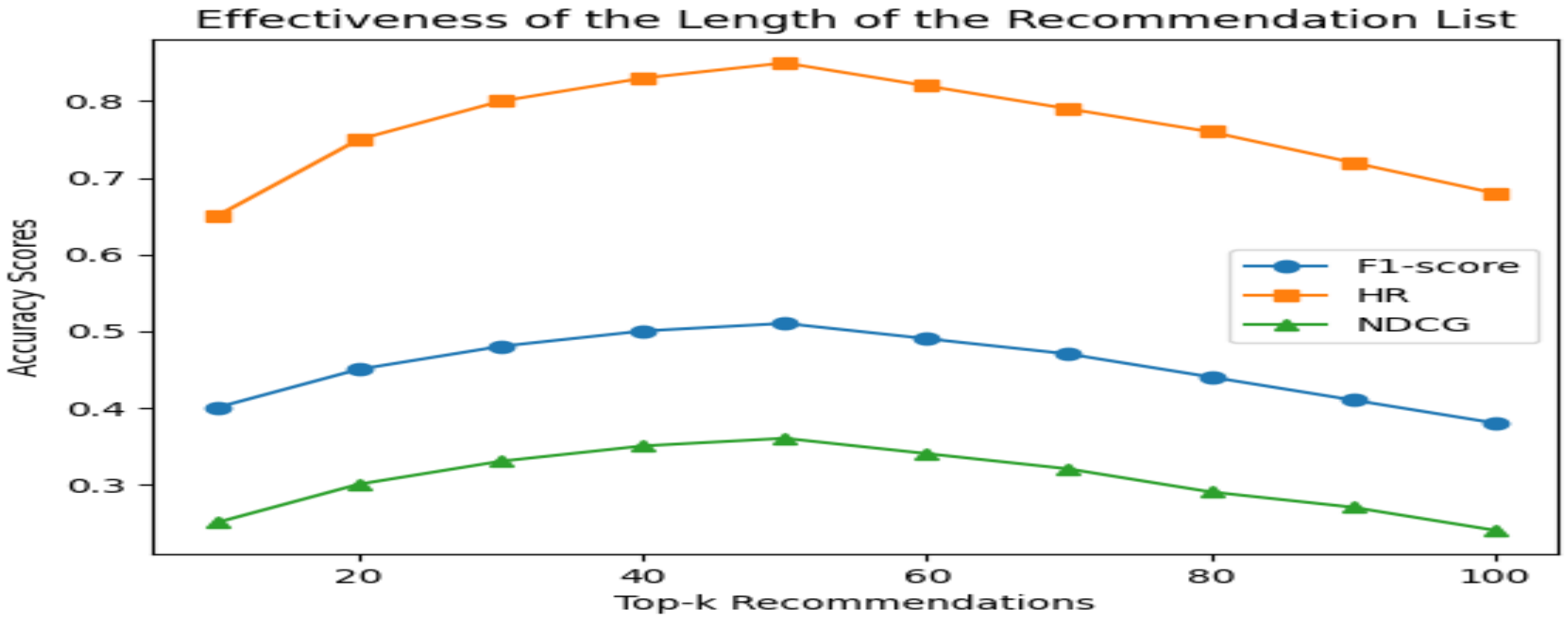

**Figure 7.** Effects of Recommendation List Length on Performance.

*5.4 Comparison of Cold-Start Approaches*

To evaluate the performance of our model in cold-start scenarios, we conducted experiments using three different approaches: collaborative filtering, content-based filtering, and a hybrid approach, these are well-known recommender system approach as reported in the literature [3,41]. For collaborative filtering, we used a matrix factorization approach to recommend items based on similar users or items. For content-based filtering, we used the textual description of the items to recommend similar items to the user. For the hybrid approach, we combined the matrix factorization and content-based filtering approaches. We evaluated the performance of these approaches in the cold-start scenario, where there is limited information about new users or items.

**Table 4.** BERT4Loc (cold-start scenario)

| Model | Precision | Recall | F1-score | HR | NDCG |
|---|---|---|---|---|---|
| Collaborative filtering | 0.38 | 0.45 | 0.41 | 0.79 | 0.35 |
| Content-based filtering | 0.29 | 0.34 | 0.31 | 0.72 | 0.27 |
| Hybrid approach | 0.47 | 0.53 | 0.49 | 0.86 | 0.42 |
| BERT4LOC | **0.78** | **0.49** | **0.60** | **0.92** | **0.71** |

As we can see from Table 4, BERT4LOC outperforms the other models in all evaluation metrics in the cold-start scenario. It has the highest recall, precision, F1-score, HR, and NDCG, indicating that it is the most effective model in recommending items for new users or items with limited information. This is probably because BERT4LOC leverages the power of Transformer-based architectures, which can handle sequential data and contextual relationships effectively. By integrating both location information and user preferences, BERT4LOC can provide more personalized and relevant recommendations, even in situations where there is limited information about the user or the item.

## 6. Discussion

Our study proposes a new approach for location recommendation, BERT4Loc, which combines the strengths of both collaborative filtering and content-based filtering techniques. Our approach shows promising results in terms of accuracy and efficiency in location recommendation.

The BERT4Loc approach has several benefits over traditional location recommendation techniques. First, it takes into account the user's context and preferences, including historical check-ins, the user's profile, and location attributes, to provide more accurate recommendations. Second, BERT4Loc is a like hybrid approach, which means it combines the strengths of collaborative filtering and content-based filtering. This approach makes it more robust and capable of handling cold-start scenarios. Finally, BERT4Loc is built on top of BERT, a state-of-the-art language model, which can learn complex patterns and relationships among user, location, and context data, resulting in more accurate recommendations.

Despite its promising results, our approach has some limitations. One limitation is that it requires a significant amount of training data to achieve high accuracy. This can be a problem for location-based services that do not have a large user base or have limited data. Another limitation is that the approach may not work well for rare or new locations that have limited data. In such cases, the model may fail to capture the patterns and relationships among the user, location, and context data, resulting in inaccurate recommendations.

There are several potential directions for future research on the BERT4Loc approach. One direction is to explore the use of more advanced deep learning models, such as graph neural networks or more foundation models, to improve the accuracy and efficiency of location recommendations. Another direction is to investigate the use of other types of data, such as social network data or sensor data, to enhance the performance of the BERT4Loc model.

Numerous other possibilities remain to be explored. We need to evaluate our model using additional datasets. A worthwhile direction is to incorporate rich POI features into the model, such as those related to the location, such as coordinates, weather, neighborhood. Another interesting direction for future work would be to incorporate more information into the user encoder of the model to enable explicit user modelling when users are logged in multiple times.

Finally, it is important to evaluate the BERT4Loc approach on real-world datasets and compare its performance with other state-of-the-art techniques to gain more insights into its strengths and weaknesses. Despite these limitations, we believe that our approach can be further improved and extended in future research to address the challenges and opportunities in location-based services.

## 7. Conclusion

The BERT architecture has been extremely successful in terms of language comprehension. We introduce a BERT for location-aware recommendations in this paper. Our model is built on top of the BERT architecture, which includes a user encoder, a POI item encoder, and a task for preference prediction and recommendation. We treat the problem of recommending POIs to a user as a sequential problem, with the task of predicting the user's next POI of interest. Extensive experimental results on real-world datasets demonstrate that our model outperforms established benchmarks. The results highlight the potential of incorporating language models in recommendation systems.

However, there are limitations to our approach, such as the dependence on large amounts of data for training and the potential for overfitting. Future directions include exploring additional features and data sources to improve model performance and evaluating the scalability of the model for larger datasets.

**Author Contributions**: Conceptualization, S.R.B., V.M. and S.R.; methodology, S.R.B, V.M., S.R.; validation, S.R.B., V.M. and S.R.; formal analysis, S.R.B., V.M., investigation, V.M.; writing—original draft preparation, S.R.B.; writing—review and editing, S.R.B., V.M. and S.R, visualization, S.R.B.; supervision, V.M.; funding acquisition, V.M.. All authors have read and agreed to the published version of the manuscript.